\documentclass[submission,copyright,creativecommons]{eptcs}
\providecommand{\event}{ICLP 2023} %

\usepackage[T1]{fontenc}

\usepackage{iftex}

\usepackage{graphicx}
\usepackage{changepage}

\usepackage[utf8]{inputenc}
\usepackage{stmaryrd}
\usepackage{fancybox}

\usepackage{algorithm}
\usepackage[noend]{algpseudocode}

\usepackage[textsize=tiny,prependcaption]{todonotes}
\newcommand{\selfnote}[1]{\todo[backgroundcolor=green!20]{#1}}
\renewcommand{\selfnote}[1]{}

\usepackage{xspace}

\usepackage[ciao]{prologrun}

\renewcommand{\ciao}{\texttt{Ciao}\xspace}
\newcommand{\ciaopp}{\texttt{CiaoPP}\xspace}

\newcommand{\lpdoc}{\texttt{LPdoc}\xspace}

\usepackage{listings}
\makeatletter
\let\old@lstKV@SwitchCases\lstKV@SwitchCases
\def\lstKV@SwitchCases#1#2#3{}
\makeatother
\usepackage{lstlinebgrd}
\makeatletter
\let\lstKV@SwitchCases\old@lstKV@SwitchCases

\lst@Key{numbers}{none}{%
    \def\lst@PlaceNumber{\lst@linebgrd}%
    \lstKV@SwitchCases{#1}%
    {none:\\%
     left:\def\lst@PlaceNumber{\llap{\normalfont
                \lst@numberstyle{\thelstnumber}\kern\lst@numbersep}\lst@linebgrd}\\%
     right:\def\lst@PlaceNumber{\rlap{\normalfont
                \kern\linewidth \kern\lst@numbersep
                \lst@numberstyle{\thelstnumber}}\lst@linebgrd}%
    }{\PackageError{Listings}{Numbers #1 unknown}\@ehc}}
\makeatother
\usepackage{textcomp}

\lstdefinestyle{MyInline}
{
  basicstyle = \relsize{-0.25}\ttfamily\color{black},
  breaklines = true,
  breakatwhitespace = true,
  upquote = true,
}
\lstdefinestyle{MySCASP}
{
  keywords = {},
  upquote = true,
  basicstyle = \ttfamily\color{PrologPredicate},
  moredelim = {**[is][\color{PrologComment}]{`}{`}},
  moredelim = {*[s][\color{black!40!PrologPredicate}]{\#pred}{.}},
  moredelim = {*[s][\color{black!40!PrologPredicate}]{\#show}{.}},
  moredelim = {*[s][\color{black!40!PrologPredicate}]{\#hide}{.}},
  moredelim = {*[s][\color{PrologVar}]{(}{)}},
  moredelim = {*[s][\color{PrologString}]{'}{'}},
  moredelim = {*[s][\color{PrologOther}]{:-}{.}},
  moredelim = {*[s][\color{red}]{/*}{*/}},
  commentstyle = \mdseries\color{PrologComment},
  morecomment=[l]\%,
  morecomment=[s]{/*}{*/},
  literate     =
  {|}{{$\mid$}}1
  {\ │}{{$\mid$}}1
  {[}{{\color{PrologOther}\small[}}1
  {]}{{\color{PrologOther}\small]}}1
  {\\$}{{\$}}1
  {&(}{{\color{PrologOther}(}}1
  {&)}{{\color{PrologOther})}}1
  {&.}{{.}}0
  {\\=}{{\char"5C=}}2
  {\\$}{{\$}}1,
}
\definecolor{codebox}{rgb}{0.996, 0.976, 0.875}

\usepackage{xcolor}
\definecolor{PrologPredicate}{RGB}{0,0,200}
\definecolor{PrologVar}      {RGB}{145,032,039}
\definecolor{PrologComment}  {RGB}{169,082,044}
\definecolor{PrologOther}    {rgb}{0.2,0.2,0.2}
\definecolor{PrologString}   {RGB}{70,100,200}

\ifpdf
  \usepackage{underscore}         %
  \usepackage[T1]{fontenc}        %
\else
  \usepackage{breakurl}           %
\fi

\title{Demonstrating (Hybrid) Active Logic Documents and the Ciao\\
  \hspace*{-3mm}Prolog Playground, and an Application to Verification
  Tutorials\hspace*{-3mm} \ $^{\small \thanks{This research has received funding from the \emph{PROCODE} Project (PID2019-108528RB-C21/MCIN/AEI/10.13039/501100011033) and the \emph{PRODIGY} Project (TED2021-132464B-I00) funded by MCIN/AEI/10.13039/501100011033/ and the European Union NextGenerationEU/ PRTR.}}$}
\newcommand{\titlerunning}{(Hybrid) Active Logic Documents, Prolog Playground,
  and an Application} %

\newcommand{\imdea}{IMDEA Software Institute, Madrid, Spain}
\newcommand{\upm}{Universidad Polit\'{e}cnica de Madrid (UPM)}
\newcommand{\evora}{NOVA LINCS / University of \'Evora, Portugal}

\author{$^{1,2}$Daniela Ferreiro, $^{1,2}$Jos\'{e} F. Morales, $^3$ Salvador Abreu and $^{1,2}$Manuel V. Hermenegildo
       \institute{$^1$ \upm\ and $^2$\imdea} 
       \institute{$^3$\evora} 
       \email{\{daniela.ferreiro,josef.morales,manuel.hermenegildo\}@imdea.org; 
       spa@uevora.pt }                   
}

\newcommand{\authorrunning}{D. Ferreiro, J.F. Morales, S. Abreu, and M.V. Hermenegildo}

\hypersetup{
  bookmarksnumbered,
  pdftitle    = {\titlerunning},
  pdfauthor   = {\authorrunning},
  pdfsubject  = {EPTCS},               %
  pdfkeywords = {Static Analysis,
    Assertions,
    Web Playground,
    Teaching Prolog,
    Ciao-Prolog,
    Logic Programming,
    Browser-based applications,
    Active Logic Documents.
  } %
}

\usepackage{hyperref}

\begin{document}
\maketitle

\begin{figure}[b]
  \vspace*{-3mm}
  \centering 
  \includegraphics[width=0.75\textwidth,clip,trim=0 0 0 1 ]{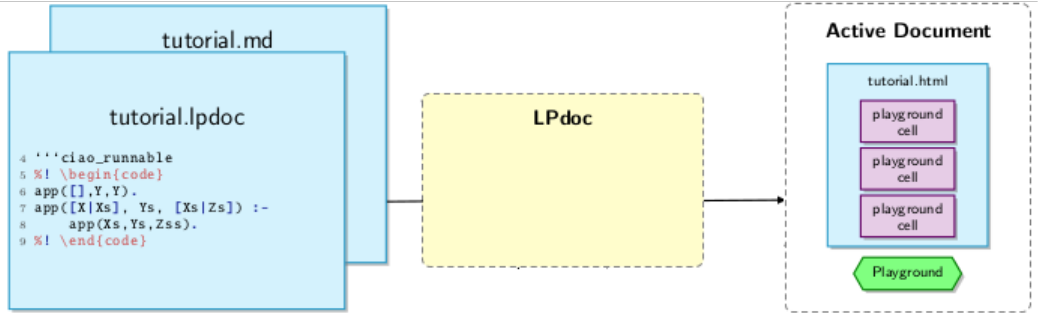}
  \vspace*{-5mm}
  \caption{Generating Active Logic Documents.}
  \label{fig:archald}
  \vspace*{-3mm}
\end{figure} 

We propose a demonstration of the Active Logic Documents (ALDs)
approach and the \ciao Playground, as well as a recent extension to
ALDs to facilitate the integration of other tools into the system for
creating Hybrid Active Logic Documents (HALD), and a concrete
application of these technologies.

\medskip

\textbf{Active Logic Documents} (ALD)~\cite{ActiveLogicDocuments-tr,ActiveLogicDocuments-PrologBook}
are web pages which incorporate embedded Prolog engines.
ALD's are generated using the \lpdoc documentation generation
system~\cite{lpdoc-reference-manual-3.0-short,lpdoc-cl2000}, extended
to be able to include embedded \ciao Playground cells in documents, so
that interaction with Prolog is possible within the documents produced
(Fig.~\ref{fig:archald}).
Documents can contain integrated environments that allow for the
processing of code blocks, including compiling, parsing, and analyzing
them.
This facilitates the creation of web-based materials with editable and
runnable educational resources such as activities, runnable examples,
programming exercises, etc.
With the ability to directly edit and evaluate code within the
document, these interactive documents can act as oracles, providing
valuable feedback and supporting self-evaluation mechanisms.
In comparison with other
tools~\cite{DBLP:journals/tplp/WielemakerRKLSC19,SimplyLogical-PrologBook,jupyter-prolog-koerner},
two fundamental aspects of the Active Logic Documents approach are
that a) all the reactive parts run locally on the user's web browser
without any dependency on a central server or a local Prolog
installation, and that b) output is generated from a single, easy to
use source that can be developed with any editor.
We argue that the ALD approach has multiple advantages from the point
of view of scalability, low maintenance cost, security, privacy, ease
of packaging and distribution, etc.\ over other approaches.
As an example,
the ALD source 
and output for a simple Prolog
exercise\footnote{\url{http://ciao-lang.org/ciao/build/doc/ciao_playground.html/factorial_peano_iso.html}}  
is shown in Fig.~\ref{fig:factorialwhole}.
Active Logic Documents are used for development of materials for
teaching logic programming~\cite{TeachingProlog-PrologBook}, embedding
runnable code and exercises in slides~\footnote{E.g., Course material
  in Computational Logic: \url{https://cliplab.org/~logalg}}, manuals,
etc., and %
in other projects, such as, for example, in the development of a
Programming course for young children (around 10 years old) within the
Year Of Prolog initiatives.

\medskip

\begin{figure}[tb]
  \centering
  \includegraphics[width=0.70\textwidth,trim=0 180 0 0,clip]{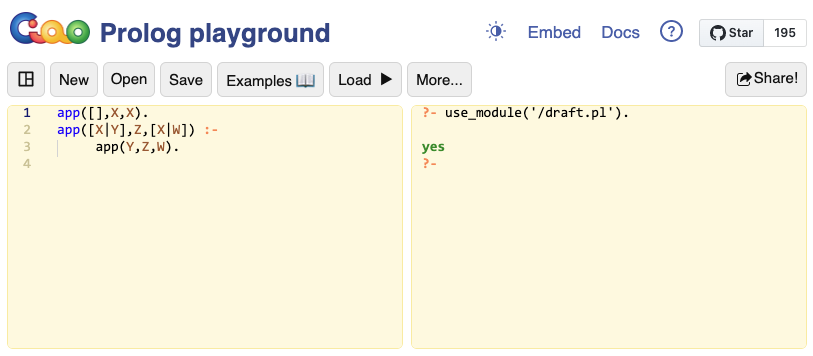}\\ [-5mm]
  \caption{The \ciao Prolog Playground}
  \label{fig:playground}
  \vspace*{-4mm}
\end{figure}

\textbf{The \ciao Prolog Playground}~\cite{ciao-playground-manual,scasp-web-gde}
is a key component %
of our approach.  In its standalone
form,\footnote{\url{https://ciao-lang.org/playground}} it allows
editing and running code locally on the user's web browser (See
Fig.~\ref{fig:playground}).  To this end, the playground uses modern
Web technology (WebAssembly and Emscripten) to run an off-the-shelf
Prolog engine and top level \emph{directly in the browser},
with access to browser-side local resources, including a full
interface with JavaScript that enables graphical input and output,
interactivity, etc. 
This engine is
contained in \texttt{Ciaowasm}, a \ciao Prolog bundle that compiles a
variant of the standard \ciao engine with all necessary \ciao bundles
and manuals. The main advantage of this general architecture over the
widely used server-based approaches such as
SWISH~\cite{DBLP:journals/tplp/WielemakerRKLSC19,SimplyLogical-PrologBook}
or Jupyter notebooks~\cite{jupyter-prolog-koerner}, is that it is easily reproducible
and significantly alleviates
maintenance effort and cost, as it essentially eliminates the
server-side infrastructure.
Previously, compilation of Prolog to
JavaScript~\cite{clp-to-js-iclp12} enabled the use of Prolog to
develop also the client side of web applications, running fully on the
browser. This functionality is also provided by, e.g., Tau
Prolog~\cite{homepageTau} and the tuProlog 
playground~\footnote{\url{https://pika-lab.gitlab.io/tuprolog/2p-kt-web}}
which are recent Prolog interpreters written in JavaScript. However, the approach
of compilation to WebAssembly and Emscripten
allows the immediate reuse of existing, well-developed engines (which
include many non trivial optimizations and features) and tooling.

\begin{figure}[tb]
  \vspace*{-5.5mm}
  \centering 
  \includegraphics[width=0.7\textwidth]{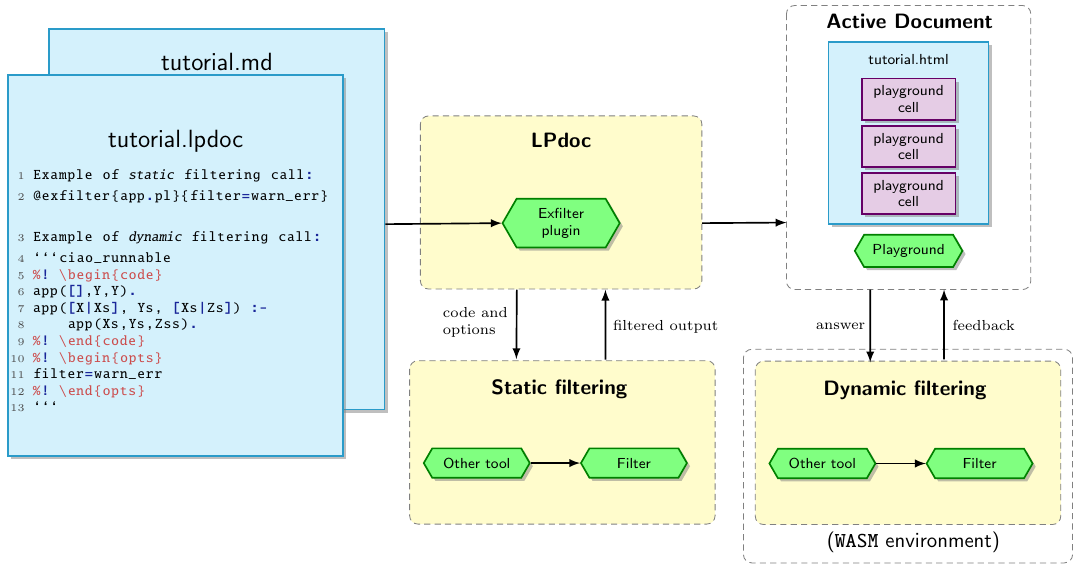}
  \vspace*{-5mm}
  \caption{Overall architecture of hybrid ALDs.}
  \label{fig:arch}
  \vspace*{-6mm}
\end{figure}

\textbf{Hybrid Active Logic Documents (HALDs).}
This addition to the ALD architecture, illustrated in
Fig.~\ref{fig:arch}, allows for the integration into ALDs of
output from other tools, that can be run either at
document generation time (to produce \emph{static} content) or during the user interactions with the
generated documents (producing the content \emph{dynamically}). 
The overall input is, as in Fig.~\ref{fig:archald}, a set of \lpdoc
source files (e.g., in \texttt{.md} markdown format or \texttt{.lpdoc}
documentation files) including code blocks, narrative text, etc.
In the \textbf{static phase}, center of Fig.~\ref{fig:arch}, \lpdoc
scans and composes all these files, processing the different elements
and inserting embedded Prolog playground instances (editors, engine
instances, query blocks, etc.) as necessary.  In addition, in the case
of a \emph{hybrid} document, one or more auxiliary tool(s) are
additionally run by \lpdoc while generating the ADL so that selected
output from such tools can be incorporated. %
We refer to
this process of projection of tool output as \textbf{filtering}. There
are a number of library filters available for this purpose, and this 
set can be extended by the user.
When \lpdoc finds filtering calls (again, center of
Fig.\ref{fig:arch}, and see also
Fig.~\ref{fig:assertioncheckingwhole}), it sends requests to the
corresponding tool (including code fragments, command line options,
queries at its top level, etc.), and then \emph{filters} the output
obtained and incorporates this projected output statically into the
document being produced.  This is very useful for example when writing
lecture notes, tutorials, manuals, etc.:  when including, e.g., the
results to a query or exercise, instead of pasting them in manually,
these results can be generated automatically from the tool being used
or documented. This ensures that the contents and format of the
outputs produced by the tool are kept automatically in sync with any
changes in the tool, which is a tedious and error-prone task without
automation.
This includes also for example obtaining from the tool the
results to compare to in student exercises.

The \textbf{dynamic phase} (right of Fig.\ref{fig:arch}) occurs once
the HTML pages produced by \lpdoc are deployed, and users have loaded
them into their browsers. When processing some user input in one of
the interactive elements of the pages --e.g., when a student issues a
solution to an exercise--, the embedded playground framework issues the
intervening filtering calls \textit{dynamically}.  This process
involves loading the program, calling the %
corresponding tool (via command line interface or interacting with its
top level), obtaining the output from this execution, and processing
it with the specified filter in order to select the relevant part
before presenting it to the user.
Filtering can be customized to accommodate various types of tasks,
such as "fill in the blanks" style exercises or to facilitate
understanding and remediation of testing and verification failures, by
displaying warning or error messages.
Thus, students can test their code rapidly and effectively and receive
useful feedback, all within the browser.
As mentioned before, the embedded Prolog is a full system with all
necessary \ciao bundles, manuals, etc, so that not just filtering but
also the tool called or any user applications can be loaded and
called to run locally on the browser, provided they are written in Prolog, or
also of course JavaScript, compiled to wasm.  
Alternatively, calls can also be
issued to tools available at some server or installed locally.
Overall, the framework greatly facilitates the whole process, saving
much coding.

\textbf{An example application of
  HALDs.} %
We will also present a concrete application
in the
generation of hybrid, interactive, web-based documentation, tutorials,
courses, etc., for a \emph{program verification tool},
\ciaopp~\cite{ciaopp-sas03-journal-scp-short,assrt-theoret-framework-lopstr99}.
The aim is assisting students in learning how to use an advanced
verification tool like \ciaopp, as well as learning more about Prolog.
The HALD approach is a great aid here %
for adding solutions to %
exercises, examples, etc.\ automatically,
while keeping the content synchronized with the
system, %
greatly reducing
the %
tracking of what documentation needs
to be changed when an update to the system is made.
\ciaopp~\cite{ciaopp-sas03-journal-scp-short,assrt-theoret-framework-lopstr99} performs
several program debugging, analysis, and source-to-source
transformation tasks %
for
Prolog programs, and also for %
other high- and low- level languages.
The output produced by \ciaopp generally contains significant amounts of information,
including transformations, static analysis information, results of
assertion checking or testing, %
counterexamples, etc. These results are typically presented as a new version of the source
file annotated with (additional) assertions. The full analysis results produced by \ciaopp
can be quite large, and cover the whole file or program. When writing the
documentation it is interesting to show only a small fraction of this information at a
time: the particular part that helps to understand the topic or step being explained.
The filters are used here to extract,
e.g., particular properties of a concrete predicate, particular
types of assertions, etc. %
Analysis information can easily be embedded in human-readable explanations.
A %
set of tutorials is available
at \url{https://ciao-lang.org/ciao/build/doc/ciaopp_tutorials.html/}
and we also refer again to Fig.~\ref{fig:assertioncheckingwhole} in
the appendix.

\smallskip
The \textbf{source code} of all components of our system is available at
\url{https://github.com/ciao-lang/}, including the build/install instructions.
When the \ciao playground is installed, all the bundles and documentation will
be run and hosted on your localhost server. 
Additional examples and instructions can be found in the
\ciao playground manual~\cite{ciao-playground-manual}~\footnote{\url{https://ciao-lang.org/ciao/build/doc/ciao_playground.html/}}.
Although we have developed and described our work for concreteness in
the context of the \ciao system, we believe the proposed mechanisms
are of general applicability and %
readily adaptable to other systems.

\bibliographystyle{eptcs}
\bibliography{extracted}

\begin{thebibliography}{10}
\providecommand{\bibitemdeclare}[2]{}
\providecommand{\surnamestart}{}
\providecommand{\surnameend}{}
\providecommand{\urlprefix}{Available at }
\providecommand{\url}[1]{\texttt{#1}}
\providecommand{\href}[2]{\texttt{#2}}
\providecommand{\urlalt}[2]{\href{#1}{#2}}
\providecommand{\doi}[1]{doi:\urlalt{https://doi.org/#1}{#1}}
\providecommand{\eprint}[1]{arXiv:\urlalt{https://arxiv.org/abs/#1}{#1}}
\providecommand{\bibinfo}[2]{#2}

\bibitemdeclare{inproceedings}{jupyter-prolog-koerner}
\bibitem{jupyter-prolog-koerner}
\bibinfo{author}{Anne \surnamestart Brecklinghaus\surnameend} \&
  \bibinfo{author}{Philipp \surnamestart Koerner\surnameend}
  (\bibinfo{year}{2022}): \emph{\bibinfo{title}{A {J}upyter Kernel for
  {P}rolog}}.
\newblock In: {\slshape \bibinfo{booktitle}{Proc. 36th Workshop on (Constraint)
  Logic Lrogramming (WLP 2022)}}, \bibinfo{series}{Lecture Notes in Informatics
  (LNI)}, \bibinfo{publisher}{Gesellschaft für Informatik, Bonn}.

\bibitemdeclare{incollection}{SimplyLogical-PrologBook}
\bibitem{SimplyLogical-PrologBook}
\bibinfo{author}{Peter \surnamestart Flach\surnameend}, \bibinfo{author}{Kacper
  \surnamestart Sokol\surnameend} \& \bibinfo{author}{Jan \surnamestart
  Wielemaker\surnameend} (\bibinfo{year}{2023}): \emph{\bibinfo{title}{{S}imply
  {L}ogical - {T}he {F}irst {T}hree {D}ecades}}.
\newblock In \bibinfo{editor}{David~S. \surnamestart Warren\surnameend},
  \bibinfo{editor}{Veronica \surnamestart Dahl\surnameend},
  \bibinfo{editor}{Thomas \surnamestart Eiter\surnameend},
  \bibinfo{editor}{Manuel \surnamestart Hermenegildo\surnameend},
  \bibinfo{editor}{Robert \surnamestart Kowalski\surnameend} \&
  \bibinfo{editor}{Francesca \surnamestart Rossi\surnameend}, editors:
  {\slshape \bibinfo{booktitle}{Prolog - The Next 50 Years}}, {\slshape
  \bibinfo{series}{LNCS}} \bibinfo{volume}{13900},
  \bibinfo{publisher}{Springer}, \doi{10.1007/978-3-031-35254-6_15}.

\bibitemdeclare{techreport}{ciao-playground-manual}
\bibitem{ciao-playground-manual}
\bibinfo{author}{G.~\surnamestart Garcia-Pradales\surnameend},
  \bibinfo{author}{J.F. \surnamestart Morales\surnameend} \&
  \bibinfo{author}{M.~V. \surnamestart Hermenegildo\surnameend}
  (\bibinfo{year}{2021}): \emph{\bibinfo{title}{{T}he {C}iao {P}layground}}.
\newblock \bibinfo{type}{Technical Report}, \bibinfo{institution}{Technical
  University of Madrid (UPM) and IMDEA Software Institute}.
\newblock
  \urlprefix\url{http://ciao-lang.org/ciao/build/doc/ciao_playground.html/ciao_playground_manual.html}.

\bibitemdeclare{inproceedings}{scasp-web-gde}
\bibitem{scasp-web-gde}
\bibinfo{author}{G.~\surnamestart Garcia-Pradales\surnameend},
  \bibinfo{author}{J.F. \surnamestart Morales\surnameend},
  \bibinfo{author}{M.~V. \surnamestart Hermenegildo\surnameend},
  \bibinfo{author}{J.~\surnamestart Arias\surnameend} \&
  \bibinfo{author}{M.~\surnamestart Carro\surnameend} (\bibinfo{year}{2022}):
  \emph{\bibinfo{title}{{A}n s({CASP}) {I}n-{B}rowser {P}layground based on
  {C}iao {P}rolog}}.
\newblock In: {\slshape \bibinfo{booktitle}{ICLP'22 Workshop on Goal-directed
  Execution of Answer Set Programs}}.

\bibitemdeclare{article}{ciaopp-sas03-journal-scp-short}
\bibitem{ciaopp-sas03-journal-scp-short}
\bibinfo{author}{M.~\surnamestart Hermenegildo\surnameend},
  \bibinfo{author}{G.~\surnamestart Puebla\surnameend},
  \bibinfo{author}{F.~\surnamestart Bueno\surnameend} \&
  \bibinfo{author}{P.~Lopez \surnamestart Garcia\surnameend}
  (\bibinfo{year}{2005}): \emph{\bibinfo{title}{{I}ntegrated {P}rogram
  {D}ebugging, {V}erification, and {O}ptimization {U}sing {A}bstract
  {I}nterpretation (and {T}he {C}iao {S}ystem {P}reprocessor)}}.
\newblock {\slshape \bibinfo{journal}{Science of Computer Programming}}
  \bibinfo{volume}{58}(\bibinfo{number}{1--2}), pp. \bibinfo{pages}{115--140},
  \doi{10.1016/j.scico.2005.02.006}.

\bibitemdeclare{inproceedings}{lpdoc-cl2000}
\bibitem{lpdoc-cl2000}
\bibinfo{author}{M.~V. \surnamestart Hermenegildo\surnameend}
  (\bibinfo{year}{2000}): \emph{\bibinfo{title}{{A} {D}ocumentation {G}enerator
  for {(C)LP} {S}ystems}}.
\newblock In: {\slshape \bibinfo{booktitle}{International Conference on
  Computational Logic, CL2000}}, {\slshape \bibinfo{series}{LNAI}}
  \bibinfo{volume}{1861}, \bibinfo{publisher}{Springer-Verlag}, pp.
  \bibinfo{pages}{1345--1361}.

\bibitemdeclare{techreport}{lpdoc-reference-manual-3.0-short}
\bibitem{lpdoc-reference-manual-3.0-short}
\bibinfo{author}{M.~V. \surnamestart Hermenegildo\surnameend} \&
  \bibinfo{author}{J.F. \surnamestart Morales\surnameend}
  (\bibinfo{year}{2011}): \emph{\bibinfo{title}{{T}he {LP}doc {D}ocumentation
  {G}enerator. {R}ef. {M}anual (V3.0)}}.
\newblock \bibinfo{type}{Technical Report}, \bibinfo{institution}{UPM}.
\newblock \bibinfo{note}{Available at \texttt{http://ciao-lang.org}}.

\bibitemdeclare{incollection}{TeachingProlog-PrologBook}
\bibitem{TeachingProlog-PrologBook}
\bibinfo{author}{M.V. \surnamestart Hermenegildo\surnameend},
  \bibinfo{author}{J.F. \surnamestart Morales\surnameend} \&
  \bibinfo{author}{P.~\surnamestart Lopez-Garcia\surnameend}
  (\bibinfo{year}{2023}): \emph{\bibinfo{title}{{S}ome {T}houghts on {H}ow to
  {T}each {P}rolog}}.
\newblock In \bibinfo{editor}{David~S. \surnamestart Warren\surnameend},
  \bibinfo{editor}{Veronica \surnamestart Dahl\surnameend},
  \bibinfo{editor}{Thomas \surnamestart Eiter\surnameend},
  \bibinfo{editor}{Manuel \surnamestart Hermenegildo\surnameend},
  \bibinfo{editor}{Robert \surnamestart Kowalski\surnameend} \&
  \bibinfo{editor}{Francesca \surnamestart Rossi\surnameend}, editors:
  {\slshape \bibinfo{booktitle}{Prolog - The Next 50 Years}}, {\slshape
  \bibinfo{series}{LNCS}} \bibinfo{volume}{13900},
  \bibinfo{publisher}{Springer}, \doi{10.1007/978-3-031-35254-6_9}.
\newblock
  \urlprefix\url{http://cliplab.org/papers/TeachingProlog-PrologBook.pdf}.

\bibitemdeclare{article}{clp-to-js-iclp12}
\bibitem{clp-to-js-iclp12}
\bibinfo{author}{J.~F. \surnamestart Morales\surnameend},
  \bibinfo{author}{R.~\surnamestart Haemmerl\'{e}\surnameend},
  \bibinfo{author}{M.~\surnamestart Carro\surnameend} \& \bibinfo{author}{M.~V.
  \surnamestart Hermenegildo\surnameend} (\bibinfo{year}{2012}):
  \emph{\bibinfo{title}{{L}ightweight compilation of {(C)LP} to
  {J}ava{S}cript}}.
\newblock {\slshape \bibinfo{journal}{Theory and Practice of Logic Programming,
  28th Int'l. Conference on Logic Programming (ICLP'12) Special Issue}}
  \bibinfo{volume}{12}(\bibinfo{number}{4-5}), pp. \bibinfo{pages}{755--773}.

\bibitemdeclare{techreport}{ActiveLogicDocuments-tr}
\bibitem{ActiveLogicDocuments-tr}
\bibinfo{author}{J.F. \surnamestart Morales\surnameend},
  \bibinfo{author}{S.~\surnamestart Abreu\surnameend},
  \bibinfo{author}{D.~\surnamestart Ferreiro\surnameend} \&
  \bibinfo{author}{M.V. \surnamestart Hermenegildo\surnameend}
  (\bibinfo{year}{2022}): \emph{\bibinfo{title}{{T}eaching {P}rolog with
  {A}ctive {L}ogic {D}ocuments}}.
\newblock \bibinfo{type}{Technical Report}, \bibinfo{institution}{Technical
  University of Madrid (UPM) and IMDEA Software Institute}.

\bibitemdeclare{incollection}{ActiveLogicDocuments-PrologBook}
\bibitem{ActiveLogicDocuments-PrologBook}
\bibinfo{author}{J.F. \surnamestart Morales\surnameend},
  \bibinfo{author}{Salvador \surnamestart Abreu\surnameend},
  \bibinfo{author}{D.~\surnamestart Ferreiro\surnameend} \&
  \bibinfo{author}{M.V. \surnamestart Hermenegildo\surnameend}
  (\bibinfo{year}{2023}): \emph{\bibinfo{title}{{T}eaching {P}rolog with
  {A}ctive {L}ogic {D}ocuments}}.
\newblock In \bibinfo{editor}{David~S. \surnamestart Warren\surnameend},
  \bibinfo{editor}{Veronica \surnamestart Dahl\surnameend},
  \bibinfo{editor}{Thomas \surnamestart Eiter\surnameend},
  \bibinfo{editor}{Manuel \surnamestart Hermenegildo\surnameend},
  \bibinfo{editor}{Robert \surnamestart Kowalski\surnameend} \&
  \bibinfo{editor}{Francesca \surnamestart Rossi\surnameend}, editors:
  {\slshape \bibinfo{booktitle}{Prolog - The Next 50 Years}}, {\slshape
  \bibinfo{series}{LNCS}} \bibinfo{volume}{13900},
  \bibinfo{publisher}{Springer}, \doi{10.1007/978-3-031-35254-6_14}.
\newblock
  \urlprefix\url{http://cliplab.org/papers/ActiveLogicDocuments-PrologBook.pdf}.

\bibitemdeclare{inproceedings}{assrt-theoret-framework-lopstr99}
\bibitem{assrt-theoret-framework-lopstr99}
\bibinfo{author}{G.~\surnamestart Puebla\surnameend},
  \bibinfo{author}{F.~\surnamestart Bueno\surnameend} \& \bibinfo{author}{M.~V.
  \surnamestart Hermenegildo\surnameend} (\bibinfo{year}{2000}):
  \emph{\bibinfo{title}{{C}ombined {S}tatic and {D}ynamic {A}ssertion-{B}ased
  {D}ebugging of {C}onstraint {L}ogic {P}rograms}}.
\newblock In: {\slshape \bibinfo{booktitle}{Logic-based Program Synthesis and
  Transformation (LOPSTR'99)}}, {\slshape \bibinfo{series}{LNCS}}
  \bibinfo{volume}{1817}, \bibinfo{publisher}{Springer-Verlag}, pp.
  \bibinfo{pages}{273--292}, \doi{10.1007/10720327_16}.

\bibitemdeclare{misc}{homepageTau}
\bibitem{homepageTau}
 (\bibinfo{year}{2021}): \emph{\bibinfo{title}{{$\tau$P}rolog --- An open
  source {P}rolog interpreter in JavaScript}}.
\newblock \bibinfo{howpublished}{\url{http://tau-prolog.org}}.
\newblock \bibinfo{note}{Last access: \today}.

\bibitemdeclare{article}{DBLP:journals/tplp/WielemakerRKLSC19}
\bibitem{DBLP:journals/tplp/WielemakerRKLSC19}
\bibinfo{author}{Jan \surnamestart Wielemaker\surnameend},
  \bibinfo{author}{Fabrizio \surnamestart Riguzzi\surnameend},
  \bibinfo{author}{Robert~A. \surnamestart Kowalski\surnameend},
  \bibinfo{author}{Torbj{\"{o}}rn \surnamestart Lager\surnameend},
  \bibinfo{author}{Fariba \surnamestart Sadri\surnameend} \&
  \bibinfo{author}{Miguel \surnamestart Calejo\surnameend}
  (\bibinfo{year}{2019}): \emph{\bibinfo{title}{Using {SWISH} to Realize
  Interactive Web-based Tutorials for Logic-based Languages}}.
\newblock {\slshape \bibinfo{journal}{Theory Pract. Log. Program.}}
  \bibinfo{volume}{19}(\bibinfo{number}{2}), pp. \bibinfo{pages}{229--261},
  \doi{10.1017/S1471068418000522}.

\end{thebibliography}

\appendix

\noindent
\begin{figure}[ht]
\centerline{{\Large \textbf{Appendix A -- ALD example}}}
\vspace*{17mm}
\begin{minipage}{0.48\linewidth}
  \includegraphics[width=\linewidth]{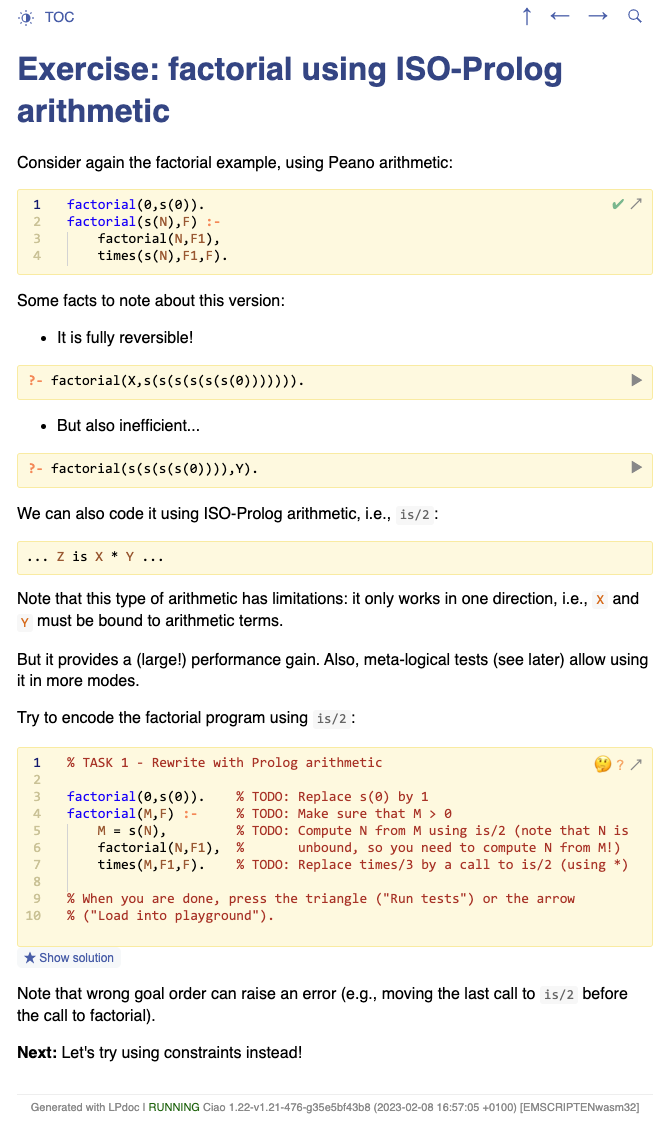}
\vspace*{15mm}
\end{minipage}
\begin{minipage}{0.48\linewidth}
    \tiny
\begin{prolog}[escapechar=@,
  linebackgroundcolor={%
    \ifnum\value{lstnumber}>0  \ifnum\value{lstnumber}<6  \color{white}   \fi \fi
    \ifnum\value{lstnumber}>4  \ifnum\value{lstnumber}<23 \color{codebox} \fi \fi
    \ifnum\value{lstnumber}>22 \ifnum\value{lstnumber}<25 \color{white}   \fi \fi
    \ifnum\value{lstnumber}>24 \ifnum\value{lstnumber}<28 \color{codebox} \fi \fi
    \ifnum\value{lstnumber}>27 \ifnum\value{lstnumber}<30 \color{white}   \fi \fi
    \ifnum\value{lstnumber}>28 \ifnum\value{lstnumber}<32 \color{codebox} \fi \fi
    \ifnum\value{lstnumber}>31 \ifnum\value{lstnumber}<34 \color{white}   \fi \fi
    \ifnum\value{lstnumber}>33 \ifnum\value{lstnumber}<37 \color{codebox} \fi \fi
    \ifnum\value{lstnumber}>36 \ifnum\value{lstnumber}<46 \color{white}   \fi \fi
    \ifnum\value{lstnumber}>45 \ifnum\value{lstnumber}<74 \color{codebox} \fi \fi
    \ifnum\value{lstnumber}>73 \ifnum\value{lstnumber}<78 \color{white}   \fi \fi
  }
  ]
\title Exercise: factorial using ISO-Prolog arithmetic 

Consider again the factorial example, using Peano
arithmetic:
```ciao\_runnable
:- module(_, _, [assertions,library(bf/bfall)]).
%! \begin{focus}
factorial(0,s(0)).
factorial(s(N),F) :-
    factorial(N,F1),
    times(s(N),F1,F).
%! \end{focus}

nat_num(0).
nat_num(s(X)) :- nat_num(X).

times(0,Y,0) :- nat_num(Y).
times(s(X),Y,Z) :- plus(W,Y,Z), times(X,Y,W).

plus(0,Y,Y) :- nat_num(Y).
plus(s(X),Y,s(Z)) :- plus(X,Y,Z).
```
Some facts to note about this version:
  - It is fully reversible!
```ciao_runnable
?- factorial(X,s(s(s(s(s(s(0))))))).
```
  - But also inefficient...
```ciao_runnable
?- factorial(s(s(s(s(0)))),Y).
```
We can also code it using ISO-Prolog arithmetic,
i.e., `is/2`:
```ciao
 ... Z is X * Y ...
```
Note that this type of arithmetic has limitations:
it only works in one direction, i.e., `X` and `Y`
must be bound to arithmetic terms.

But it provides a (large!) performance gain.  Also,
meta-logical tests (see later) allow using it in more
modes.

Try to encode the factorial program using `is/2`:
```ciao_runnable
:- module(_, _, [assertions]).
:- test factorial(5, B) => (B = 120) + (not_fails).
:- test factorial(0, 0) + fails.
:- test factorial(-1,B) + fails.
%! \begin{hint}
% TASK 1 - Rewrite with Prolog arithmetic 

factorial(0,s(0)).    % TODO: Replace s(0) by 1
factorial(M,F) :-     % TODO: Make sure that M > 0
    M = s(N),         % TODO: Compute N from M using is/2
                      % (note that N is unbound, 
    factorial(N,F1),  % so you need to compute N from M!)
    times(M,F1,F).    % TODO: Replace times/3 by a call
                      % to is/2 (using *)

% When you are done, press the triangle ("Run tests") or
% the arrow ("Load into playground").
%! \end{hint}
%! \begin{solution}
factorial(0,1).
factorial(N,F) :-
    N > 0,
    N1 is N-1,
    factorial(N1,F1),
    F is F1*N.
%! \end{solution}
```
Note that wrong goal order can raise an error (e.g.,
moving the last call to `is/2` before the call to
factorial).
**Next:** Let's try using constraints instead!
\end{prolog}
\vspace*{-2mm}
\end{minipage}
\caption{The full source and \lpdoc output for the Active Logic Document for a simple factorial exercise.}
\label{fig:factorialwhole}
\end{figure}

\noindent
\begin{figure}[ht]
\centerline{{\Large \textbf{Appendix B -- HALD example}}}
\vspace*{17mm}
\begin{minipage}{0.48\linewidth}
  \centerline{\ciaoinline{Output (exercise.html)}}
  \vspace*{10mm}
  \includegraphics[width=\linewidth]{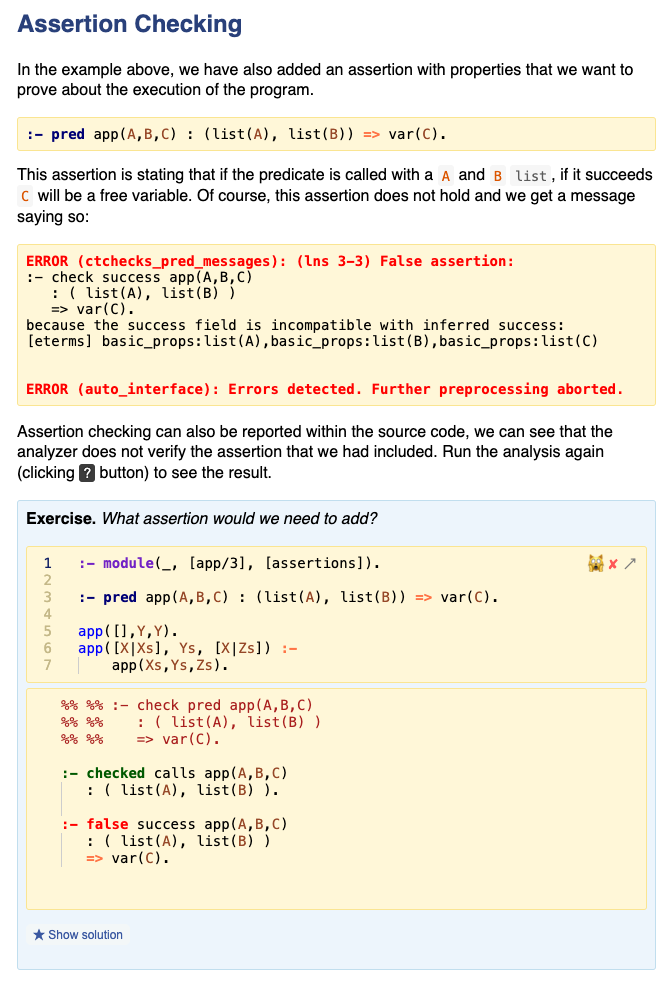}
\vspace*{15mm}
\end{minipage}
\begin{minipage}{0.48\linewidth}
  \vspace*{-24.5mm}
  \centerline{\ciaoinline{Input (exercise.md)}}
  \vspace*{11mm}
    \tiny
\begin{prolog}[
  linebackgroundcolor={
    \ifnum\value{lstnumber}>0  \ifnum\value{lstnumber}<14  \color{white}   \fi \fi
    \ifnum\value{lstnumber}>13  \ifnum\value{lstnumber}<15 \color{codebox} \fi \fi
    \ifnum\value{lstnumber}>14  \ifnum\value{lstnumber}<22 \color{white}   \fi \fi
    \ifnum\value{lstnumber}>21 \ifnum\value{lstnumber}<45 \color{codebox} \fi \fi
  }
  ]
# Assertion Checking

In the example above, we have also added an assertion
with properties that we want to prove about the execution
of the program.
```ciao
:- pred app(A,B,C) : (list(A), list(B)) => var(C).
```
This assertion is stating that if the predicate is called
with a A and B list, if it succeeds C will be a free
variable. Of course, this assertion does not hold and we
get a message saying so:

@exfilter{app_assrt_false.pl}{V,filter=warn_error}

Assertion checking can also be reported within the source
code, we can see that the analyzer does not verify the
assertion that we had included. Run the analysis again
(clicking ? button) to see the result.

Exercise. What assertion would we need to add?
```ciao_runnable
%! \begin{code}
:- module(_, [app/3], [assertions]).

:- pred app(A,B,C) : (list(A), list(B)) => var(C).

app([],Y,Y).
app([X|Xs], Ys, [X|Zs]) :-
    app(Xs,Ys,Zs).  
%! \end{code}
%! \begin{opts}
solution=verify_assert
%! \end{opts}  
%! \begin{solution}
:- module(_, [app/3], [assertions]).

:- pred app(A,B,C) : (list(A), list(B)) => list(C).

app([],Y,Y). 
app([X|Xs], Ys, [X|Zs]) :-
    app(Xs,Ys,Zs).   
%! \end{solution}
```
\end{prolog}
\end{minipage}

\vspace*{-10mm}

\caption{The full source and \lpdoc output for a Hybrid Active Logic
  Document containing a simple assertion checking exercise. Note the
  use of a filtering command in the source to call \ciaopp during the
  static phase and select certain outputs that are incorporated in the
  page (the box with the error messages). Also, an interactive
  exercise is embedded and filtering is used again (this time
  dynamically) to present a selected part of the analysis output for  
  the program entered by the student and to compare it to the expected
  results.}
\label{fig:assertioncheckingwhole}
\end{figure}

\end{document}